\documentclass[a4paper,twocolumn,floatfix]{revtex4}
\usepackage{graphicx}

\begin{document}

\title{Electronic transport through electron-doped Metal-Phthalocyanine Materials}

\author{M. F. Craciun$^{1}$}
\author{S. Rogge$^{1}$}
\author{M. J. L. den Boer$^{1}$}
\author{S. Margadonna$^{2}$}
\author{K. Prassides$^{3}$}
\author{Y. Iwasa$^{4, 5}$}
\author{A. F. Morpurgo$^{1}$}

\affiliation{$^{1}$Kavli Institute of Nanoscience, Delft
University of Technology, Lorentzweg~1, 2628\,CJ Delft, The
Netherlands}

\affiliation{$^{2}$School of Chemistry, University of Edinburgh,
West Mains Road, Edinburgh EH9 3JJ, UK}

\affiliation{$^{3}$Department of Chemistry, University of Durham,
South Road, Durham DH1 3LE, UK}

\affiliation{$^{4}$Institute for Materials Research, Tohoku
University, Katahira 211, Aoba-ku, Sendai 980-8577, Japan}

\affiliation{$^{5}$CREST, Japan Science and Technology Corporation
Kawaguchi 332-0012, Japan}

\begin{abstract}
We report an insulator-metal-insulator transition in films of five
metal phthalocyanines (MPc) doped with alkali atoms. Electrical
conduction measurements demonstrate that increasing the alkali
concentration results in the formation of a metallic state for all
systems. Upon further doping, the films reenter the insulating
state. Structural and Raman spectroscopy studies reveal the
formation of new crystalline phases upon doping and are consistent
with the phenomena originating from charge transfer between the
intercalated alkali atoms and MPc, in a similar fashion to what
has been so far observed only in C$_{60}$. Due to the presence of
a molecular spin, large exchange energy, and a two-fold orbital
degeneracy in MPc, our findings are of interest in the study of
controllable magnetism in molecular materials and in the
investigation of new, recently predicted electronic phases.

\end{abstract}

\maketitle

Metal phthalocyanines (MPc's) form a large class of molecules
consisting of a stable $\pi$-conjugated macrocyclic ligand bonded
to a central metallic atom. They are very well known for their
electronic properties, which are both of applied and fundamental
interest \cite{MPcbook}. In their pure form, MPc materials behave
as semiconductors. Electrical conduction can be induced through
doping, usually by oxidizing the ligands and creating an open
shell, which results in the introduction of holes in the materials
\cite{MPcreview1,MPcreview2}. Depending on the specific MPc
molecule and the degree of oxidation, a rich variety of molecular
conductors have been developed in this way. Surprisingly, in spite
of the rich behavior of hole-doped MPc compounds and even though
it has been shown that multiple reduction processes enable a large
accumulation of electrons on different orbitals of many MPc's
\cite{Clack,Taube}, the conducting, electronic and magnetic
properties of electron-doped MPc materials have remained so far
vastly unexplored.

In this paper we demonstrate experimentally that it is possible to
induce metallic conduction in electron-doped MPc materials and
control their electronic properties, through intercalation with
alkali atoms.  We will first discuss the transport properties of
five different compounds - CuPc, NiPc, CoPc, FePc, and MnPc - that
we have investigated as a function of potassium concentration and
temperature. For all these systems, insulating in the pristine
state, we found that the electrical conductivity can be increased
to a value in excess of $\sigma$=100 S$\cdot$cm$^{-1}$ upon
potassium intercalation. In this state, the conductivity of all
compounds remains high at cryogenic temperatures, indicating the
occurrence of metallic behavior. Increasing further the potassium
concentration brings all MPc's back into an insulating state. We
will then focus on one of these molecular systems, CuPc, to
address the electronic and structural properties of this compound
in more detail. We discuss scanning tunnelling spectroscopy
experiments that show the presence of a finite density of states
in the metallic state and of a gap in the insulating states, thus
confirming at the local level the occurrence of the
insulator-metal-insulator transition. We also present structural
investigations demonstrating the formation of intercalated phases
and Raman spectroscopy measurements, which provide an independent
microscopic demonstration of charge transfer between K and MPc.

Many aspects of our results bear a clear resemblance to the
behavior of C$_{60}$ (or, more in general, the fullerides), the
only other molecular system in which a metallic temperature
dependence of the conductivity induced by alkali doping has been
measured in the past \cite{Haddon,Hebard1}. As opposed to
C$_{60}$, MPc's enable the tuning of many relevant parameters,
such as the degeneracy of the low-energy molecular levels, the
spin in the molecular ground state, the shape of the molecular
orbitals \cite{orbitals} and facile chemical tuning. In view of
the well-established and rich electronic behavior of hole-doped
MPc's \cite{MPcreview1,MPcreview2} and of alkali-doped fullerides
\cite{Forro,Hebard2,Takenobu}, we expect that controlling these
molecular parameters will lead to the emergence of a rich physical
phenomenology. For instance, the ability to tune the molecular
spin and the large values of the exchange energy in MPc's open the
possibility to control magnetic phenomena, known to occur in
different MPc compounds \cite{Magn1,Magn2,Magn3}. In addition, the
two-fold degeneracy of the relevant orbitals of many MPc's has led
to the prediction of new electronic phases around half-filling
\cite{Tosatti}. The overall picture emerging from our experiments,
together with these considerations, clearly indicates the richness
of alkali-doped metal phthalocyanines as a model class of
compounds for the controlled investigation of electron-doped
molecular solids.

\begin{figure}[h]
  \centering
  \includegraphics[width= 1\columnwidth]{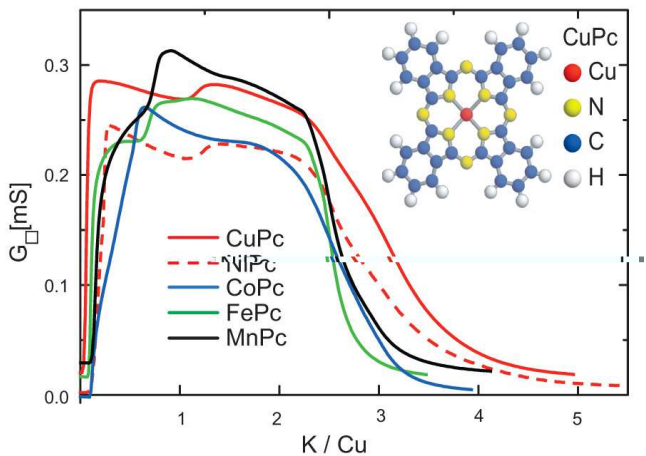}
  \caption { The square conductance measured at room
temperature as a function of potassium concentration on films of
different MPc's. The inset shows the structure of a CuPc
molecule.}
  \label{fig:MPcdoping}
\end{figure}

The investigation of the transport properties of alkali-doped
MPc's has been performed on thin films, thermally evaporated on
the surface of a silicon-on-insulator (SOI) substrate. Film
deposition, doping, scanning probe characterization, and transport
measurements have been carried out \textit{in-situ} in a single
ultra-high vacuum (UHV) system. Figure \ref{fig:MPcdoping} shows
the conductance of films of the different MPc's as a function of
doping concentration.

For all the materials, the conductance first increases with
potassium concentration up to a high value that is comparable for
the different molecules, it remains high in a broad range of
concentrations, and it eventually decreases to the level observed
for the pristine material. The high conductivity state (optimally
doped state) occurs in a broad interval up to approximately three
potassium atoms per molecule, corresponding to a much higher
carrier density than what is typically achieved in holed-doped MPc
compounds \cite{MPcreview1,MPcreview2}. From the value of the
maximum conductance we find that for all MPc's the highest
electron mobility is close to 1 cm$^{2}$/Vs, comparable to that of
C$_{60}$ \cite{Forro}. We established the high reproducibility of
this behavior by growing, doping, and measuring over 200 films. In
the course of these experiments we also observed experimentally
robust differences in the doping dependence of the conductance for
the different MPc's (visible in Fig. \ref{fig:MPcdoping}). These
differences are interesting in that they systematically correlate
to the known properties of the MPc molecular orbitals and suggest
that electrons can be added to orbitals that are centered either
on the ligands or on the metal atoms, depending on the specific
molecule. This behavior \cite{MPc-JACS} expected on the basis of
existing calculations for individual MPc molecules
\cite{orbitals}, is different from the case of hole-doped MPc
compounds, in which the holes typically occupy the ligand orbitals
only.

\begin{figure}[ht]
  \centering
  \includegraphics[width=1\columnwidth]{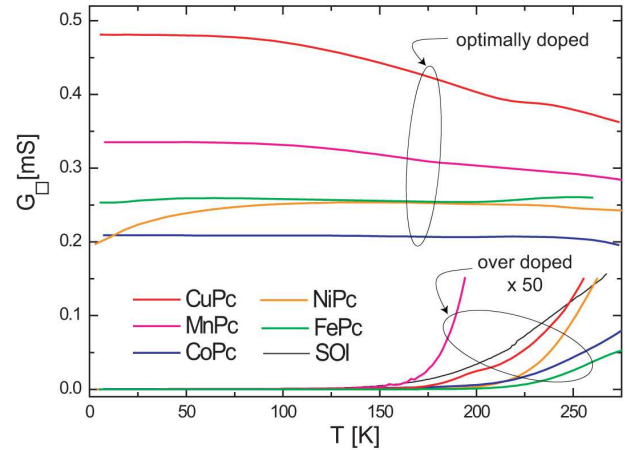}
\caption{Temperature dependence of the square conductance for
films of the different MPc's measured into the highly conductive
(optimally doped) state and in the overdoped state (rescaled by a
factor of 50). The (insulating) temperature dependence of the SOI
substrate is also shown.}
  \label{fig:GT}
\end{figure}

To investigate the nature of electrical conduction in the MPc
films, we have measured the temperature dependence of the
conductivity for different doping levels (Fig. \ref{fig:GT}). For
the undoped films, the conductance is decreasing with decreasing
temperature as expected for insulating materials. When the films
are doped into the highly conductive state, for all MPc's the
conductance remains high when lowering the temperature and the
films are conducting down to low temperature, demonstrating the
occurrence of metallic conduction. In the overdoped state the
conductance again decreases very rapidly with decreasing
temperature. This evolution of the temperature dependence of the
conductance unambiguously reveals that an
insulator-metal-insulator transition occurs in all five MPc's upon
doping with K.

\begin{figure}[ht]
  \centering
  \includegraphics[width=1\columnwidth]{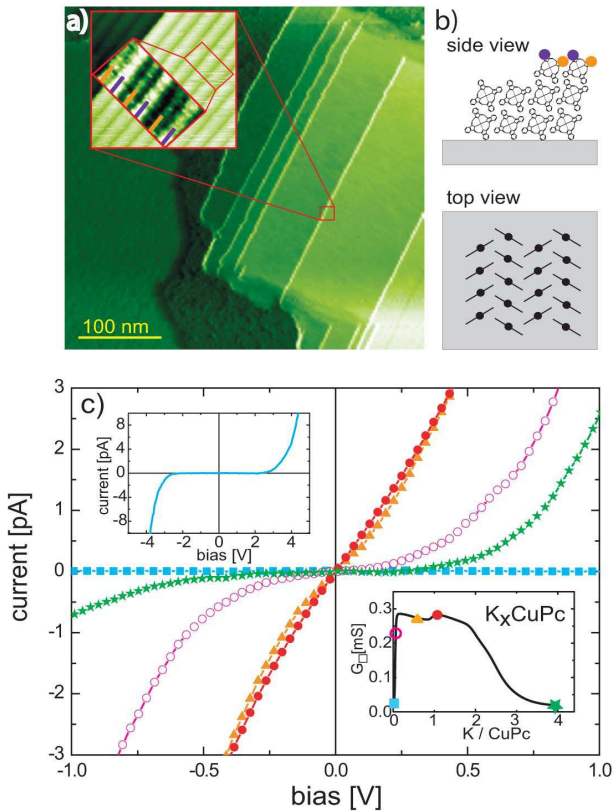}
\caption{(a) STM images of a CuPc film showing molecular (first
inset) and submolecular structure (second inset). (b) The expected
molecular structure of the CuPc films. The purple and yellow
colored phenyl rings corresponds to the features marked with the
same colors in the inset of a). (c) Scanning tunneling
spectroscopy of K-CuPc at different doping levels. The symbols on
the conductance versus doping curve in the lower-right inset
approximately indicate the doping level at which the $I-V$ curves
of the corresponding color have been measured.}
  \label{fig:STM}
\end{figure}

In order to investigate the microscopic nature of the
insulator-metal-insulator transition, we have performed scanning
tunnelling spectroscopy on highly ordered CuPc films, a few
monolayer thick, before and after K-doping. To make this possible,
films with large domains were grown on 2$\times$1-reconstructed
Si(001) surface. High resolution scanning tunneling microscopy
(STM) images of our films are shown in Fig.\ref{fig:STM}a. The
film morphology consists of a fully closed first monolayer and
highly ordered islands typical of the Stranski-Krastanov growth
mode. Zooming in on a molecular terrace (first inset in
Fig.\ref{fig:STM}a) reveals that the molecules grow in ordered
columns which lie flat on the Si(001) surface \cite{aip,growth1}
as shown in the schematic diagram of the molecular structure of
Fig.\ref{fig:STM}b. Further zooming in on the molecular columns
(second inset in Fig.\ref{fig:STM}a) shows that each column
exhibit features corresponding to the uppermost phenyl rings of a
CuPc molecule.

STS measurements on thin, undoped CuPc films shows a very large
apparent gap of several volts (Fig.\ref{fig:STM}, upper inset).
This is because undoped films are highly resistive and a
considerable fraction of the voltage applied between the substrate
and the STM tip drops across the CuPc layers (\textit{i.e.}, the
gap in the {\it I-V} curves is not the HOMO-LUMO gap of CuPc).
Upon doping, the magnitude of the gap rapidly decreases
(Fig.\ref{fig:STM}). When the doping level corresponds to a state
of high conductivity, the tunneling gap vanishes and a finite
slope is observed in the tunneling $I-V$ curves around zero bias
voltage, which indicates a finite density of states in the doped
film. Upon further doping of the films, a large gap in the
tunneling curve is observed again. Therefore, the STS data fully
support the conclusion drawn from the temperature dependence of
the conductance: at low and high doping level the CuPc film is
insulating, in the intermediate region it is metallic (see
Fig.\ref{fig:GT}). Other MPc's on which we have performed similar,
but less systematic, STS experiments show a similar behavior.

The behavior of K-doped CuPc films suggests a progressive filling
of one or more molecular orbitals with electrons transferred from
potassium into the molecular film. The complete filling is mainly
supported by the fact that after reaching the metallic state, the
conductance of the film decreases to the level of the pristine
material. Note that this observation excludes the possibility that
the observed conduction is an experimental artifact, such as, for
instance, the formation of a potassium surface layer. Further, we
have observed that the doping process is reversible: overdoped
K-CuPc films can be de-doped by exposure to oxygen, which removes
electrons from the film by oxidizing K. This excludes the
occurrence of irreversible chemical reactions between K and the
MPc's.\\

With the exception of the fullerenes, this is the first time that
a molecular system has been shown to exhibit an
insulator-metal-insulator transition upon doping with alkali
atoms. Previous work on different molecules had shown that charge
transfer from the alkali atoms does occur and that, in certain
cases, a finite density of states can be induced
\cite{PES-KCuPc,PES1,PES2,pentacene,Iwasaki}. However, a
low-temperature metallic conduction had never been observed in
these systems, which typically exhibited thermally activated
decrease of conductivity with lowering temperature \cite{Iwasaki}.
Interestingly, such a behavior is what we observe when we measure
CuPc films grown on SiO$_2$ substrates. The difference between
films grown on Si and on SiO$_2$ originates from structural
disorder, which for films grown on SiO$_2$ has been proven to be
much larger than for films grown on Si \cite{growth1,growth2}. The
large influence of disorder on transport properties is consistent
with our observation that the highest carrier mobility in MPc
films deposited on SiO$_2$ is only $\mu \ \approx \ 10^{-3}$
cm$^2$/Vs, \textit{i.e.} $10^3$ times smaller than the mobility
found in films deposited on a Si surface.\\

Having established the similarity of the phenomenology upon alkali
intercalation of MPc's and C$_{60}$ thin films, we then started to
explore the structural properties of the materials. In particular,
whereas in pristine C$_{60}$, there is well-defined interstitial
space to accommodate the alkali ions in the parent crystalline
structure, it is not immediately obvious how this might be
achieved in pristine MPc's without major structural modifications.
Since it is technically impossible for us to undertake {\it
in-situ} X-ray diffraction (XRD) and Raman measurements in the
same UHV system used for the investigation of transport
properties, we have performed these experiments on polycrystalline
powdered samples.

\begin{figure}[ht]
  \centering
  \includegraphics[width=1\columnwidth]{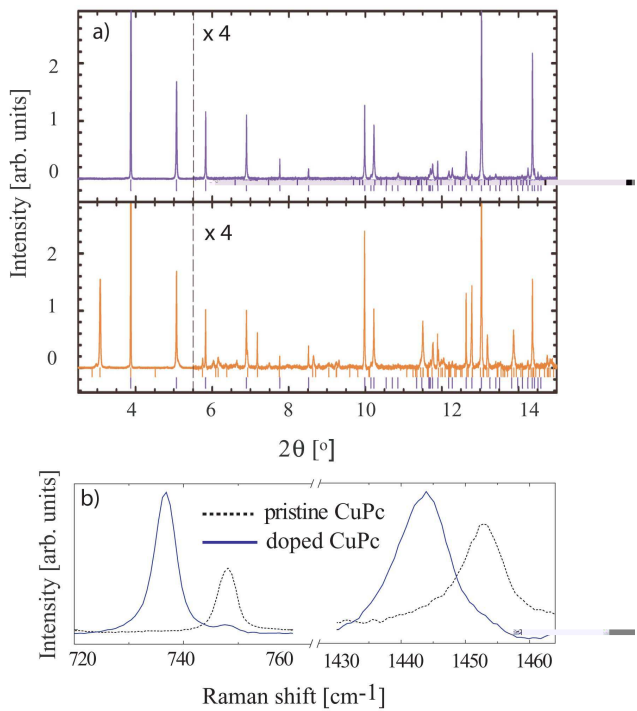}
\caption{(a) Synchrotron X-ray powder diffraction profiles
($\lambda$ = 0.85001 \AA) of pristine CuPc (top) and K-doped CuPc
(bottom panel) samples at 300 K. The profiles have been expanded
for clarity by a factor of 4 at Bragg angles larger than
5.5$^{o}$. In each case, the tick marks show the reflection
positions (blue: CuPc, red: K[CuPc]). Panel (b) shows selected
regions of the Raman spectra of the same pristine and K-doped CuPc
samples, which illustrates the shift of the vibrational lines.}
  \label{fig:XRD}
\end{figure}

Figure \ref{fig:XRD}a shows the high-resolution synchrotron XRD
powder patterns of pristine ($\beta$-phase \cite {betaCuPc}) and
K-doped CuPc. The measured diffraction profile of the material
after K-doping comprises peaks arising from a residual fraction of
the pristine compound together with additional reflections due to
the formation of a new crystalline phase. The latter can be
indexed with a primitive monoclinic unit cell with lattice
constants {\it a} = 25.2311(4) \AA, {\it b} = 4.4049(3) \AA, {\it
c} = 21.6272(5) \AA, and $\beta$ = 93.924(4)$^{o}$ and an expanded
unit cell volume relative to the pristine copper phthalocyanine by
12 \AA$^{3}$/CuPc unit. These findings demonstrate the existence
of at least one stable intercalated phase in the potassium-CuPc
phase field. Work is in progress towards complete structural
determination and to search for additional intercalated phases at
different doping levels.

We have also performed Raman spectroscopy studies on the same
samples used for the structural characterization and compared the
results for the doped and pristine material (see
Fig.\ref{fig:XRD}). The undoped sample shows features, assigned to
the C-N and C-H bonds of the CuPc molecule in agreement with
previous studies \cite {betaCuPc}. Many of these lines are also
found unshifted in the doped sample, consistent with the presence
of pristine material. However, as shown in Fig. \ref{fig:XRD}, the
doped material also exhibits new vibrational lines shifted towards
lower wavenumbers. The values found for the doping-induced Raman
shifts range from 7.6 cm$^{-1}$ to 18.6 cm$^{-1}$ for different
lines. This shift confirms the occurrence of charge transfer from
the potassium atoms to the CuPc molecules.

In conclusion, we have performed a systematic investigation of the
electronic and structural properties of electron-doped metal
phthalocyanine compounds, through the intercalation of potassium
atoms. As compared to known hole-doped materials based on the same
molecules, electron doping gives access to a broader range of
electron density. This has enabled us to reveal an
insulator-metal-insulator transition in the conductivity of the
molecular materials with increasing doping, similar to what had
been so far only observed in C$_{60}$. The combination of the
large density range accessible by electron doping and the various
molecular properties of metal phthalocyanines, \textit{i.e.},
controllable spin, two-fold orbital degeneracy, orbital symmetry,
and large exchange energy, make alkali-doped metal phthalocyanine
compounds unique for the study of new electronic phenomena,
ranging from controllable magnetism to the occurrence of novel,
recently predicted, electronic phases \cite{Tosatti}.\\

We thank J. van den Brink and E. Tosatti for useful discussions,
Y. Taguchi and T. Miyake for experimental help, A.N. Fitch for
help with the diffraction experiments, the ESRF for provision of
beamtime and are grateful to AMOLF for the RBS analysis of our
samples. This work was financially supported by the stichting FOM,
the Netherlands Organization for Scientific Research (MFC), the
Royal Dutch Academy of Sciences KNAW (SR), the NWO
Vernieuwingsimpuls 2000 program (AFM), the Royal Society (Dorothy
Hodgkin Research Fellowship to SM) and the Daiwa Foundation
(KP).\\

\textbf{\textit{Experimental}}

\textit{Preparation of alkali-doped Metal-Phthalocyanines thin
films :} All the steps of our investigations have been carried out
in the UHV system with a base pressure $< 5 \times
10^{-11}$\,mbar. This prevents the occurrence of degradation of
the doped films over a period of days. Prior to the film
deposition, as-purchased compounds were heated in UHV at a
temperature just below their sublimation temperature for several
days, to remove contaminants. The films were thermally evaporated
from a Knudsen cell onto a hydrogen (wet) passivated silicon (001)
surface of a silicon-on-insulator (SOI) wafer, consisting of a top
2$\mu$m silicon layer electrically insulated from the Si substrate
by 1$\mu$m-thick SiO$_{2}$ buried layer. The use of SOI as
substrate permits to grow high-quality MPc films, which is only
possible on a crystalline Si surface \cite{growth1}, while
avoiding a large parallel electrical conduction through the entire
Si substrate. The typical film thickness was 20 nm. For the
scanning probe experiments, films with large domains, a few
monolayer thick, were grown at 500K on a 2$\times$1-reconstructed
Si(001) surface. Alkali doping was achieved by exposing the film
to a constant flux of K atoms generated by a current-heated getter
source. In order to determine the potassium concentration in the
film, we have performed an elemental analysis for several doping
levels using ex-situ RBS for CuPc. We have then used the K-CuPc
data to scale the concentration of the other molecular films as a
function of potassium exposure time and film thickness (note that
this method is affected by a relatively large uncertainty in
K/MPc, which we estimate to be approximately of the order of one
at a high doping level).\\

\textit{Preparation of polycrystalline powdered samples:}
Potassium-doped CuPc was synthesized in bulk form by a direct
reaction of K vapor and CuPc powder in sealed glass tubes, heated
at 300$^{o}$C for three weeks. Prior to this procedure,
as-purchased CuPc powder was purified in a thermal gradient by
vacuum sublimation. All sample manipulations were undertaken in an
Ar-atmosphere glove box with an oxygen concentration $\approx$ 1
ppm. After the reaction, the as-synthesized material was
introduced into glass capillaries and sealed under Ar gas.
Synchrotron X-ray powder diffraction patterns of pristine and
K-doped CuPc were recorded on the ID31 beamline at the European
Synchrotron Radiation Facility (ESRF, Grenoble, France) at ambient
temperature and $\lambda$= 0.85001 $\AA$.


\begin{thebibliography}{99}
\bibitem{MPcbook} N. B. McKeon, \textit{Phthalocyanine materials}, Cambridge University Press, (New York
1998).

\bibitem{MPcreview1} T. Inabe, and H. Tajima, \textit{Chem. Rev.} \textbf{104}, 5503
(2004).

\bibitem{MPcreview2} T. J. Marks, \textit{Angew. Chem.} \textbf{29}, 857
(1990).

\bibitem{Clack} D. W. Clack, and J. R. Yandle, \textit{Inorg. Chem.} \textbf{11}, 1738
(1975).

\bibitem{Taube} R. Taube, and H. Drevs,  \textit{Angew. Chem. Internat. Edit.} \textbf{6}, 358
(1967).

\bibitem {Haddon} R. C. Haddon, A. F. Hebard, M. J. Rosseinsky, D. W. Murphy, S. J. Duclos,
K. B. Lyons, B. Miller, J. M. Rosamilia, R. M. Fleming, A. R.
Kortan, S. H. Glarum, A. V. Makhija, A. J. Muller, R. H. Eick, S.
M. Zahurak, R. Tycko, G. Dabbagh, and F. A. Thiel, \textit{Nature}
\textbf{350}, 320 (1991).

\bibitem{Hebard1} G. P. Kochanski, A. F. Hebard, R. C. Haddon, and A. T. Fiory,
\textit{Science}, \textbf{255}, 184 (1992).

\bibitem{orbitals} M. S. Liao, and S. Scheiner, \textit{J. Chem. Phys.} \textbf{114}, 9780
(2001).


\bibitem{Forro} L. Forro, and L. Mihaly, \textit{Rep. Prog. Phys.} \textbf{64}, 649 (2001).


\bibitem{Hebard2} A. F. Hebard, M. J. Rosseinsky, R. C. Haddon, D. W. Murphy, S. H. Glarum,
T. T. M. Plastra, A. P. Ramirez, and A. R. Kortan, \textit{Nature}
\textbf{350}, 600 (1991).


\bibitem{Takenobu} T. Takenobu, T. Muro, Y. Iwasa, and T. Mitani, \textit{Phys. Rev. Lett.}
\textbf{85}, 381 (2000).


\bibitem{Magn1} M. Evangelisti, J. Bartolome, L. J. De Jongh, and G. Filoti, \textit{Phys. Rev. B}
\textbf{66}, 144410 (2002).


\bibitem{Magn2} K. Awaga, and Y. Maruyama, \textit{Phys. Rev. B} \textbf{44}, 2589 (1991).


\bibitem{Magn3} M. Y. Ogawa, S. M. Palmer, and K. Liou, \textit{Phys. Rev. B} \textbf{39}, 10682 (1989).


\bibitem{Tosatti} E. Tosatti, M. Fabrizio, J. Tobik, and G.E. Santoro \textit{Phys. Rev.
Lett.} \textbf{93}, 117002 (2004).


\bibitem{growth1} M. Nakamura, and H. Tokumoto, \textit{Surf. Sci.} \textbf{398},
143 (1998).


\bibitem{MPc-JACS} M. F. Craciun, S. Rogge, and A. F. Morpurgo, \textit{J. Am. Chem. Soc.}
\textbf{127}, 12210 (2005).


\bibitem{aip} M. F. Craciun, S. Rogge, D. A. Wismeijer, M. J. L. den Boer, T. M. Klapwijk, and A. F. Morpurgo,
 \textit{AIP Conf. Proc.} \textbf{696}, 489 (2003).


\bibitem{PES-KCuPc} T. Schwieger, M. Knupfer, W. Gao, and A. Kahn, \textit{Appl. Phys. Lett.}
\textbf{83}, 500 (2003).


\bibitem{PES1} L. Yan, N. J. Watkins, S. Zorba, Y. Gao, and C. W. Tang, \textit{Appl.
Phys. Lett.} \textbf{79}, 4148 (2001).


\bibitem{PES2} T. Schwieger, H. Peisert, M. S. Golden, M. Knupfer, and J. Fink,
\textit{Phys. Rev. B} \textbf{66}, 155207 (2002).


\bibitem{pentacene} T. Minakata, M. Ozaki, and H. Imai, \textit{J. Appl. Phys.} \textbf{74}, 1079 (1993).


\bibitem{Iwasaki} K. Iwasaki, K. Umishita, M. Sakata, and S. Hino, \textit{Synth. Metals}
\textbf{121}, 1395 (2001).


\bibitem{LeMoigne} J. Le Moigne, and R. Even, \textit{J. Chem. Phys.} \textbf{83}, 6472 (1985).


\bibitem{growth2} M. Nakamura, Y. Morita, Y. Mori, A. Ishitani, and H. Tokumoto,
\textit{J. Vac. Sci. Technol. B} \textbf{14}, 1109 (1996).


\bibitem{betaCuPc} R. Prabakaran, R. Kesavamoorthy, G. L. N. Reddy, F. P. Xavier,
\textrm{Phys. Stat. Sol. B} \textbf{229}, 1175 (2002).


\end{thebibliography}
\end{document}